\documentclass{iau-JDSS2}
\usepackage{graphicx}

\title[short title of paper] 
{The radio-infrared correlation in galaxies }

\author[short author list]   
{F.S. Tabatabaei, R. Beck, \and E. Berkhuijsen}

\affiliation{Max-Planck-Institut f\"ur Radioastronomie, Auf dem H\"ugel 69,
53121 Bonn, Germany \break email:tabataba@mpifr-bonn.mpg.de\\[\affilskip]
}

\pubyear{2009}
\volume{Volume 15}  
\pagerange{119--126}
\date{?? and in revised form ??}
\jname{Highlights of Astronomy, Volume 15}
\editors{Corbett, ed.}
\begin{document}

\maketitle



The radio--infrared correlation holds within galaxies down to scales of
about 50\,pc (\cite{Hughes_etal_06}, \cite{Tabatabaei_1_07}). It was explained as a direct and linear relationship between star formation and IR emission.  However, one fact making the IR-star formation linkage less obvious is that the IR emission consists of at least two emission components, cold dust and warm dust. The cold dust emission may not be directly linked to the young stellar population. 
Furthermore, understanding the origin of the radio--IR correlation requires to discriminate between the two main components of the radio continuum emission, free-free and synchrotron emission. Although cosmic ray electrons originates also from the star forming regions  (supernovae remnants; final episodes of massive stars), the synchrotron--IR correlation may not be as tight as thermal--IR correlation locally, as a result of convection and diffusion of the cosmic ray electrons from their place of birth. The magnetic field distribution may further modify the correlation.

We present a multi-scale study of the correlation of IR with both the thermal and non-thermal (synchrotron) components of the radio continuum emission from the nearby galaxies M33 and M31. 
Using the Spitzer MIPS IR data at 70\,$\mu$m and 160\,$\mu$m, we derive extinction maps which are used to correct H$\alpha$ and to separate the radio thermal and nonthermal components at 20\,cm, following \cite{Tabatabaei_3_07}. Our scale-by-scale analysis in M33 using wavelet functions showed that the correlation is almost perfect on all scales between the radio thermal emission and the warm dust emission.  The thermal radio--cold dust in M31 is not as tight as that in M33 on scales smaller than 2\,kpc, indicating different dust heating sources: young massive stars in M33 but the interstellar radiation field in M31. 
The synchrotron--IR correlation is better in M33 than in M31 locally, however, it is better in M31 than in M33 globally. \cite{Tabatabaei_08} showed that M33 is dominated by strong turbulent magnetic field in star forming regions. In M31, however, \cite{Fletcher_04} found a stronger large scale regular magnetic fields than the small scale turbulent field.  Therefore, the different scale-dependency of the synchrotron--IR correlation in M33 and M31 can be explained by their different magnetic fields and hence the propagation of cosmic ray electrons. Regarding that the star formation rate per area in M33 is about 10 times larger than that in M31 (Tabatabaei et al. in prep.), we conclude that  
the magnetic fields and cosmic rays enhanced in star forming regions can cause a good correlation between the synchrotron and IR emission locally. This phenomena should be visible in  late-type galaxies with on-going star formation and pronounced in starburst galaxies.


\begin{thebibliography}{}

\bibitem[Fletcher  \etal\ (2004)]{Fletcher_04}
     {Fletcher, A., Berkhuijsen, E.~M., Beck, R., Shukurov, A.} 2004,
     \textit{A\&A}  414, 53
\bibitem[Hughes  \etal\ 2006]{Hughes_etal_06}
     {Hughes, A., Wong, T., Ekers, R. \etal\ } 2006,
     \textit{MNRAS}  370, 363
\bibitem[Tabatabaei  \etal\ 2007a]{Tabatabaei_1_07}
     {Tabatabaei, F.S., Beck, R., Krause, M. \etal\ } 2007a,
     \textit{A\&A}  466, 509
\bibitem[Tabatabaei  \etal\ (2007b)]{Tabatabaei_3_07}
     {Tabatabaei, F.S., Beck, R., Kr\"ugel, E. \etal\ } 2007b,
     \textit{A\&A}  475, 133
\bibitem[Tabatabaei  \etal\ (2008)]{Tabatabaei_08}
     {Tabatabaei, F.S., Krause, M., Fletcher, A., Beck, R.} 2008,
     \textit{A\&A}  490, 1005

  
\end{thebibliography}
\end{document}